# Magnetism of Ta Dichalcogenide Monolayers Tuned by Strain and Hydrogenation

Priyanka Manchanda,[1] Vinit Sharma,[2] Hongbin Yu,[3] D. J. Sellmyer,[1] and Ralph Skomski[1]

[1]*Department of Physics and Astronomy and Nebraska Center for Materials and Nanoscience,
University of Nebraska, Lincoln, NE 68588, United States.*

[2]*Department of Materials Science and Engineering and Institute of Materials Science,
University of Connecticut, Storrs, CT 06269, United States.*

[3]*School of Electrical, Computer, and Energy Engineering,
Arizona State University, Tempe, AZ 85287, United States*

**Abstract**

The effects of strain and hydrogenation on the electronic and magnetic properties of monolayers of Ta based dichalcogenides (Ta$X_2$; $X$ = S, Se, Te) are investigated using density-functional theory. We predict a complex scenario of strain-dependent magnetic phase transitions involving paramagnetic, ferromagnetic, and modulated antiferromagnetic states. Covering one of the two chalcogenide surfaces with hydrogen switches the antiferromagnetic/nonmagnetic Ta$X_2$ monolayers to a semiconductor. Our research opens new pathways towards the manipulation of magnetic properties for future optoelectronics and spintronics applications.

PACS: 72.90.+y; 75.10.Lp; 75.50.Ee; 75.70.-i

*Introduction.* — Two-dimensional transition-metal dichalcogenides (2D TMDs) combine many advantages of graphene and conventional thin-film materials and have recently attracted much attention as potential materials for electronics, photovoltaic cells, sensors, catalysis, thermal applications, and energy storage [1-7]. Examples are flexible electronic devices, beyond-graphene nanoelectronics such as thin-film transistors [7], and transition-metal disulfide thin films in water electrolysis for hydrogen production [3]. Compared to graphene, which needs large strain to open a band gap [8, 9] and where applications exploit rather than change its basic characteristics [6], the electronic properties of 2D TMDs are very versatile. Among these features are direct band gaps in the visible wavelength range, excellent room-temperature electron mobility, mechanical properties, and easy chemical functionalization [1-5]. The exploration of these new two-dimensional materials is an important aspect of current research in nanoelectronics and other areas, as exemplified by efforts to approach the ultimate limits of thin-film transistor technology [7].

Thin films of TMDs such as $MoS_2$, $WS_2$, $MoSe_2$, $MoTe_2$, $TaSe_2$, and $NbSe_2$ exhibit excellent "peel-off" characteristics and can be easily synthesized experimentally, using methods such as chemical, liquid, and micromechanical exfoliation techniques, and chemical vapor deposition [10-15]. A related advantage of the films is their mechanical robustness, and experiment shows that TMD thin films can sustain strain of up to about 11% [16]. The properties of pristine 2D TMDs vary from semiconducting ($MoS_2$, $WS_2$) to nonmagnetic metallic ($NbSe_2$, $TaS_2$) and magnetic ($VS_2$, $VSe_2$) [17-20].

Very recently, it has become clear that 2D TMDs exhibit striking physical-property changes due to strain [16, 21-29], chemical functionalization [30-34], doping with transition metals [35], and external electric fields [36]. For instance, it has been demonstrates experimentally that strain modulates the bandgap of monolayers and bilayers in $MoS_2$ [21]. First-principle calculations using density-functional theory (DFT) play an important role in the prediction and investigation of the properties of these materials. Examples are the strong effect of uniaxial and isotropic strains on the electronic properties of early transition-metal dichalcogenide $MX_2$ monolayers ($M$ = Sc, Ti, Zr, Hf, Ta, Cr; $X$ = S, Se, Te) [22], strain-induced bandgap changes in $MoS_2$ [23], and semiconductor-to-metal transitions in $SnSe_2$ [24].

An emerging topic in 2D TMD research is the investigation of *magnetic* degrees of freedom. Magnetic effect are important by themselves, due to potential applications in spin electronics, and because they are coupled with other physical phenomena, such as electrical conductivity. This research is exemplified by calculations of the magnetic moment for different chemical compositions [22] and of strain-induced or -modified magnetism in $MoS_2$ [25, 26] and $VX_2$ ($X$ = S, Se) [27]. Some of the publsihed information is contradictory, such as the reports of strain-dependent ferromagnetism [28] and antiferromagnetism [29] in $NbX_2$ ($X$ = S, Se) monolayers.

Chemical functionalization is another approach to modify magnetic properties. Electronic-structure calculations suggest that adsorbed hydrogen molecules dissociate at the TMD surface, each hydrogen forming a bond with a chalcogen and providing an extra electron to the system [31]. In $VX_2$ monolayers ($X$ = S, Se, Te), nonmagnetic metallic, antiferromagnetic and semiconducting states have been predicted for different hydrogen coverages [32]. Some combined effects have also been studied, such as ferromagnetism in $MoS_2$ caused by strain and hydrogenation [33]. Experimentally, adsorbed fluorine has been shown very recently to create a small magnetic moment of 0.06 emu/g in $MoS_2$ nanosheets [34].

The focus of the Letter is on monolayers of the tantalum dichalcogenides $TaS_2$, $TaSe_2$, and $TaTe_2$. We use first-principles calculations to explore the strain and hydrogenation effects on the electronic and magnetic properties of this intriguing class of TMD materials and predict a complex scenario of magnetic and nonmagnetic phase transitions. From the viewpoint of electron transport, the transitions involve metallic and "semimetallic" states, as well as direct and indirect semiconductivity, whereas the magnetic phases include paramagnets, ferromagnets, and modulated antiferromagnets.

*Computational Details.* — Our density-functional theory (DFT) calculations are performed using the projector augmented wave (PAW) method [37], as implemented in the Vienna *ab-initio* simulation package (VASP) [38, 39]. Within the generalized gradient approximation (GGA), we employ Perdew-Burke-Ernzerhof (PBE) exchange and correlation functionals [40]. The kinetic energy cutoff of plane-wave expansion is taken as 520 eV. All the geomeric structures are fully relaxed until the force on each atom is less than 0.002 eV/Å, and the energy-convergence criterion is $1\times10^{-6}$ eV. For the electronic-structure calculations, a $23 \times 23 \times 1$ $k$-point grid is used. To investigate the spin structure, we construct a $4 \times 4 \times 1$ supercell, and a sufficently large vacuum is used in the vertical direction to avoid the interaction between neighboring supercells. Figures 1(a) and (e) show the top and side views of the monolayers, respectively. The big red arrows in (a) arrows show the positive isotropic strain to which the monolayers are subjected.

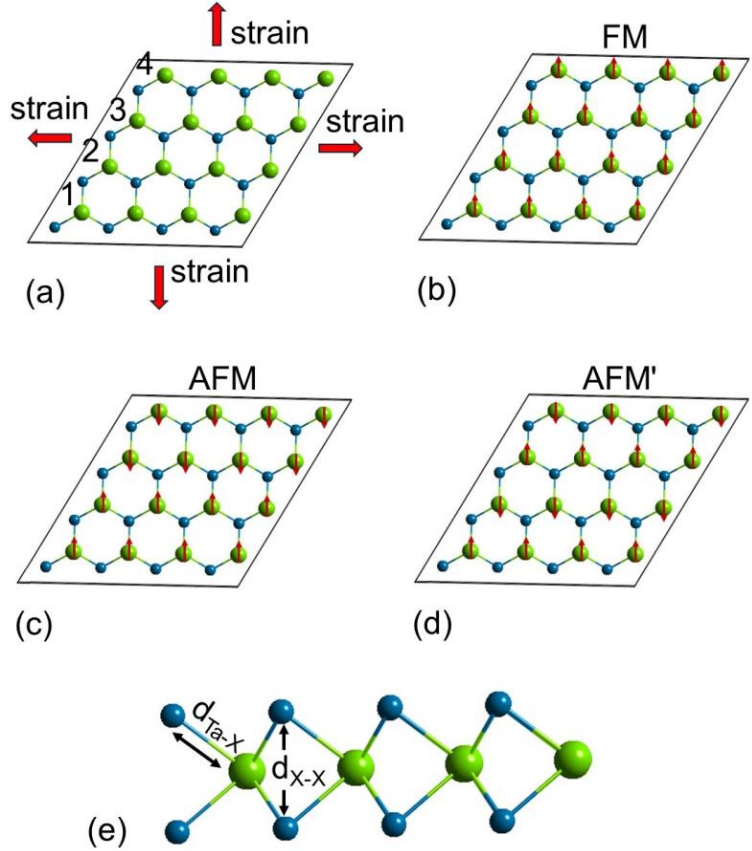

**Figure 1**. Tantalum dichalcogenide monolayers: (a) top view of $4 \times 4$ Ta$X_2$ supercell ($X$ = S, Se, Te) subjected to isotropic strain, (b) ferromagnetic (FM) spin configuration, (c) AFM spin configuration, with ↑ electrons in rows 1 and 2 and ↓ electrons in rows 3 and 4, (d) AFM′ spin configuration, with ↑ electrons in rows 1 and 3 and ↓ electrons in 2 and 4, and (e) side view of Ta$X_2$ supercell. Green and blue colors are used for Ta and $X$ atoms, respectively.

*Results.* — The structural relaxation is described in the Supplement. To study the magnetic order, we have considered four different spin configurations, namely paramagnetic (PM), ferromagnetic (FM), and two antiferromagnetic (AFM and AFM′) spin structures, the AFM′ structure as previously [29] suggested. Figures 1(a-d) show these spin configurations. In unstrained $TaTe_2$, the AFM state has the lowest energy, whereas unstrained $TaS_2$ and $TaSe_2$ are paramagnetic.

Figure 2(a) displays the strain dependence of the energies of the magnetically ordered spin configurations of $TaTe_2$ relative to the paramagnetic energy $E_{PM}$. In $TaTe_2$, the AFM state remains stable up to an isotropic strain of 2%. Above 2%, the magnetic order is ferromagnetic. The PM and AFM′ have the same energy, meaning that the AFM' state degenerates into the PM state with zero Ta atomic moments.

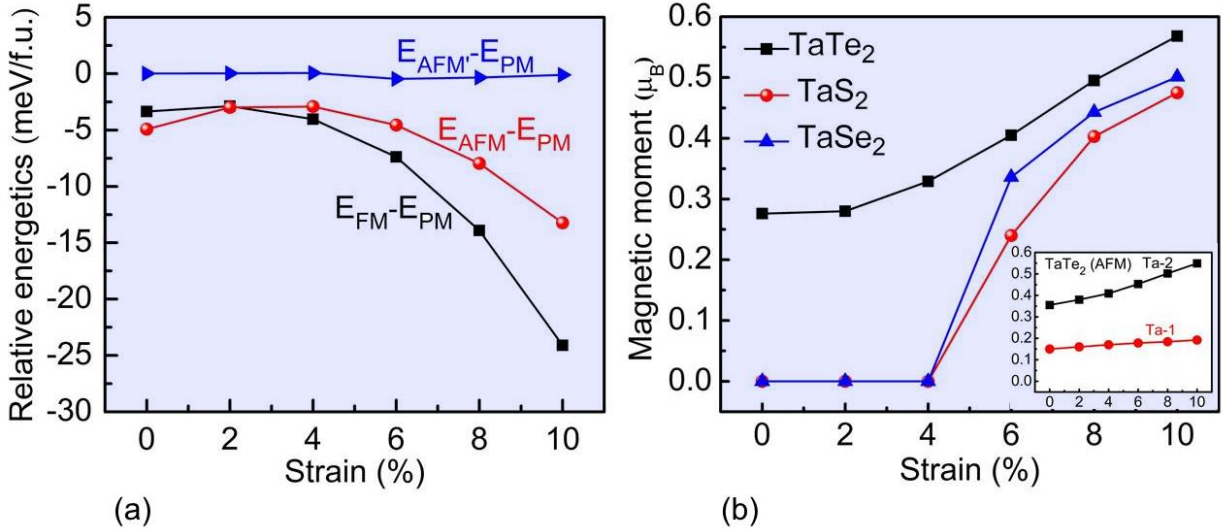

**Figure 2.** Spin structure and magnetization of Ta$X_2$ monolayers: (a) energies of the FM, AFM, and AFM' states of $TaTe_2$ relative to the PM state and (b) atomic moment, measured in $\mu_B$ per Ta atom. The inset shows the magnetic moment of the Ta-1 and Ta-2 atoms in the AFM spin configuration of Fig. 1(a).

Figure 2(b) shows the strain dependence of the atomic magnetic moments in the monolayers. In $TaTe_2$, the atomic moments corresponds to the AFM state for small strain and to the FM state for strains larger than 2%. In $TaS_2$ and $TaSe_2$, all nonzero moments are FM. In all cas-es, the atomic magnetic moments increase with increasing strain. For example, in TaTe$_2$, strains of 4% and 10% yield ferromagnetic moments of 0.28 $\mu_B$ and 0.57 $\mu_B$ per Ta atom, respectively. The inset in Fig. 2(b) plots the magnetic moments of the Ta atoms in the AFM spin configuration, Ta-1 and Ta-2 corresponding to rows 1 and 2 in Fig. 1(a), respectively. The magnetic moments per Ta atom, 0.15 $\mu_B$ (Ta-1) and 0.36 $\mu_B$ (Ta-2) in the unstrained TaTe$_2$ monolayer, are different, in spite of the structural equivalence of the two sites. This modulation, not shown in the schematic spin structure of Fig. 1(c), indicates that the spin structure in the Ta dichalcogenide thin films is very complex.

In contrast to TaTe$_2$, unstrained TaS$_2$ and TaSe$_2$ monolayers remain paramagnetic below about 6% strain. At about 6% strain, both TaS$_2$ and TaSe$_2$ become ferromagnetic with respective magnetic moments of 0.24 $\mu_B$ and 0.34 $\mu_B$ per Ta atom. The magnetic moment mainly arises from Ta atoms and the contribution of S, Se, and Te to the total magnetic moment is negligible. The moment increases with strain, and at 10%, it reaches 0.48 $\mu_B$ and 0.50 $\mu_B$ for TaS$_2$ and TaSe$_2$, respectively.

To trace the origin of the magnetic moment, we have investigated the spin-polarized density of states (DOS) of the Ta 5$d$ states. Figure 3 shows the 5$d$ DOS of TaS$_2$ and TaSe$_2$ for unstrained monolayers and for a strain of 6%. We see that the magnetic moment arises from Ta 5$d$ states near the Fermi level, whose spin degeneracy is lifted by the strain. The character of these states is predominantly of the $|m| = 2$ type, comprising the $d_{x^2-y^2}$ and $d_{xz}$ orbitals, which lie in the film plane. In contrast to graphene, there is also some admixture of $|m| = 1$ character at the Fermi level, especially $d_{xz}$, because the chalcogen atoms are located slightly above and below the Ta plane, as shown in Fig. 1(e).

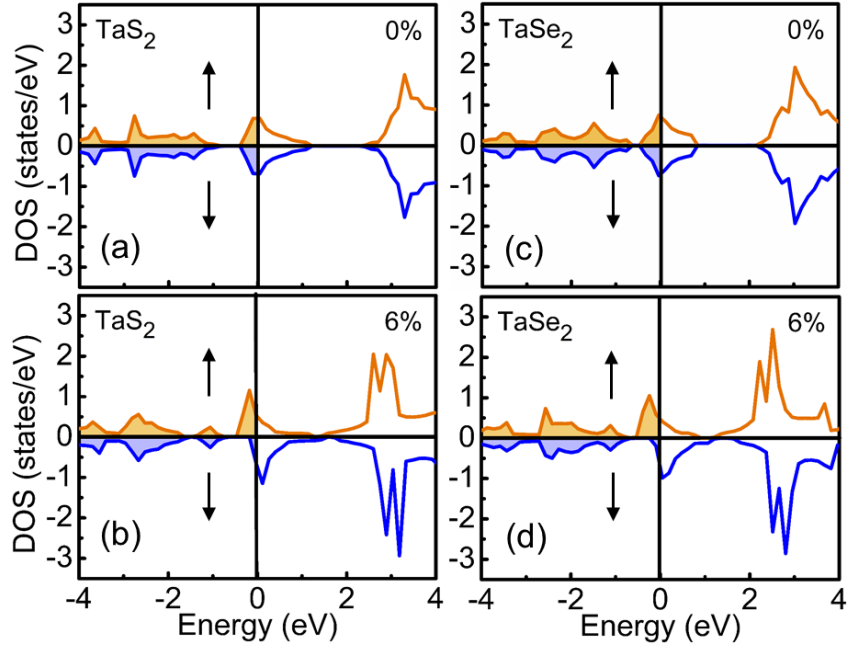

**Figure 3.** Spin-polarized partial 5$d$-density of states (DOS) of unstrained and 6% strained monolayers: (a-b) $TaS_2$ and (c-d) $TaSe_2$.

Due to its transition from antiferromagnetism to ferromagnetism, the DOS of $TaTe_2$ requires a separate consideration. Figure 4 (a-b) shows the spin-polarized local $d$-DOS of the Ta-1 and Ta-2 atoms in the unstrained AFM state. The exchange splitting of Ta-2 is larger than that of Ta-1, which corresponds to the different moments described above.

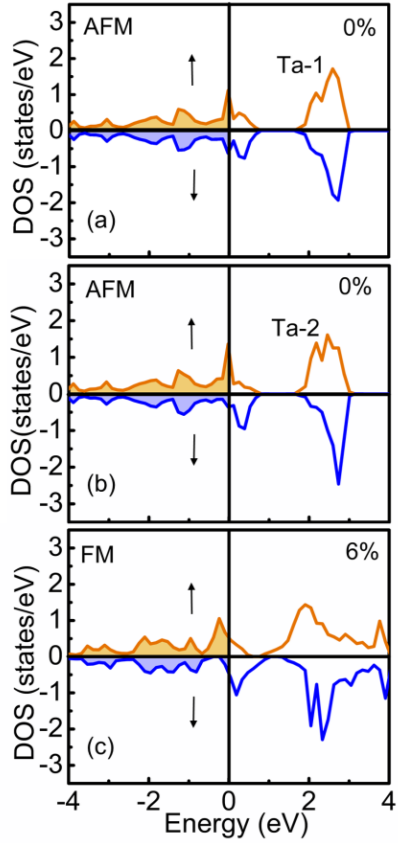

**Figure 4.** Spin-polarized partial 5$d$-density of states (DOS) of monolayers of TaSe$_2$: (a-b) unstrained AFM state and (c) FM state with 6% strain.

As mentioned in the introduction, hydrogen adsorption drastically affects the properties of some dichalcogenide thin films. We consider the electronic properties of hydrogenated-Ta$X_2$ with one surface covered by hydrogen atoms (Ta$X_2$-1H). Figure 5(a) shows a perspective view of the corresponding structure, where we follow previous research [31, 33] by assuming that the most stable hydrogen location is on top of the $X$ atoms. Our calculations show that the optimized lattice constants of Ta$X_2$-1H are expanded by 1.5% to 3.5% compared to Ta$X_2$, reaching 3.39 Å, 3.48 Å, and 3.82 Å for TaS$_2$-1H, TaSe$_2$-1H, and TaTe$_2$-1H, respectively. For S-H, Se-H and Te-H, the respective bond lengths are 1.37 Å, 1.51 Å, and 1.71 Å.

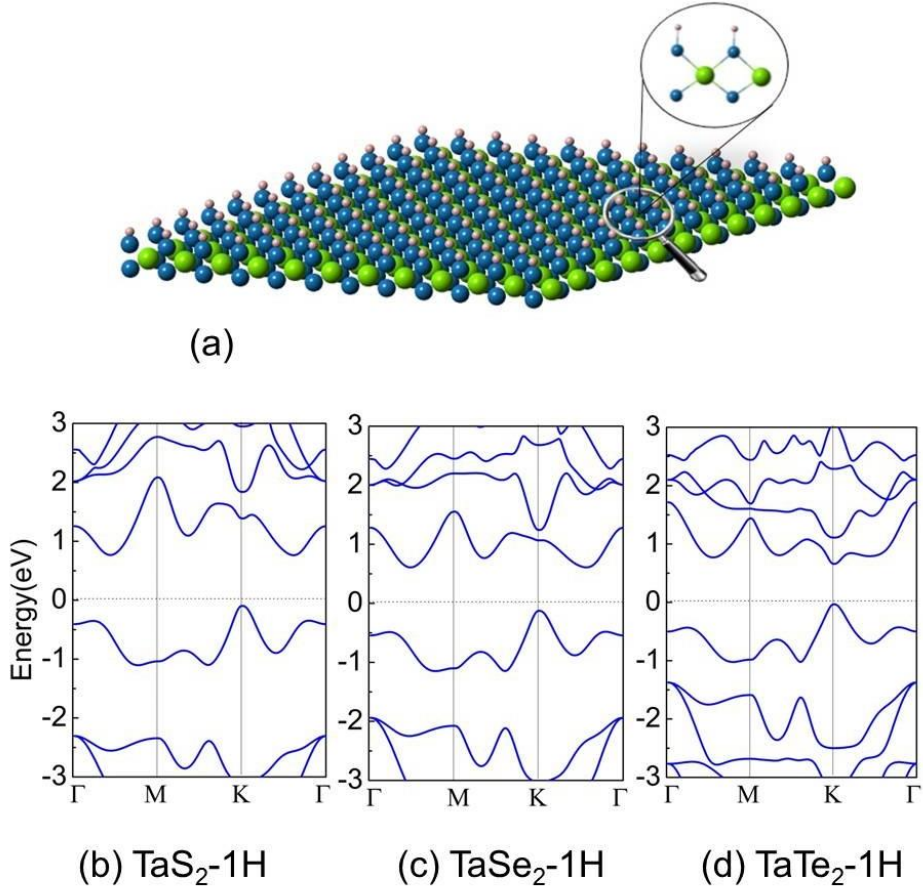

**Figure 5.** Hydrogenated Ta$X_2$ monolayers: (a) monolayer with one side fully covered with hydrogen atoms (Green: Ta, blue: $X$, gray: H), (b) band structure of TaS$_2$-1H, (c) band structure of TaSe$_2$-1H, and (d) band structure of TaTe$_2$-1H.

The hydrogenation strongly affects the electronic properties of the monolayers. Figures 5(b-d) show the calculated band structures of TaS$_2$-1H, TaSe$_2$-1H, and TaTe$_2$-1H. In contrast to the hydrogen-free monolayers, which exhibit both ↑ and ↓ states at the Fermi level (Figs. 3-4), there are no states at the Fermi level of the hydrogenated systems. This means that the hydrogenated Ta monolayers are semiconducting. From Fig. 5 we see that TaS$_2$-1H, and TaSe$_2$-1H have indirect band gaps of about 0.75 eV and 0.67 eV, respectively, whereas TaTe$_2$-1H exhibits a direct band gap of 0.62 eV. The transition from antiferromagnetic or paramagnetic metallicity to semiconducting behavior occurs because the hydrogen donates electrons that fill the empty Ta 5$d$-states just above the Fermi level, which is similar to the situation in V$X_2$ monolayers [32].

Figure 5 shows that hydrogenation causes the top of the occupied band, located at the $K$ point, to be very close to the Fermi level. For incomplete hydrogen coverage, where the Fermi level is somewhat lower, band-filling arguments suggest a zero temperature conductivity reminiscent of semimetals. However, the situation is actually more complicated, because the hole carriers move in a two-dimensional environment and are therefore subject to weak localization [41] near hydrogen vacancies.

*Discussion and Conclusions.* — Our research shows that TaX$_2$ is a particularly interesting and versatile 2D TMD system, especially with respect to strain and hydrogenation. It is instructive to compare the present findings with earlier research on *MX*$_2$ thin films. Strain-induced magnetism in TMD monolayers has been ascribed to the competition between covalent-bond and ionic-bond interactions [27, 28]. The present paper suggest a somewhat different picture, where the monolayers are solids in which the Stoner criterion favors ferromagnetism for large interatomic distances. In a broad sense, the Stoner criterion is related to ionicity [42], but there is one important aspect where the two explanations differ: the ionicity changes very little as solids undergo the Stoner transition from Pauli paramagnetism (PM) and ferromagnetism (FM), and the band structure becomes the main consideration.

Similar arguments apply to the occurrence of antiferromagnetic (AFM) order. Antiferromagnetism in *unstrained* TaTe$_2$ has been explained as a superexchange mechanism [32, 29], but as in the FM case, the AFM can also be explained in terms of the band structure, as itinerant antiferromagnetism mediated by chalcogen atoms. In fact, a previous calculation [22], also dealing with unstrained TaTe$_2$, suggests that the monolayer is actually *ferromagnetic* rather than antiferromagnetic. This finding is obtained by comparing the FM energy with the AFM′ energy, while the AFM state was not considered. Our calculations safely exclude an FM ground state and show that the ground state is more complicated than simple antiferromagnetism.

In summary, we have investigated materials beyond graphene and used density-functional calculations to study the electronic and magnetic properties of strained and hydrogenated TaS$_2$, TaSe$_2$, and TaTe$_2$ monolayers. Isotropic strain and hydrogenation yield a variety of phase transitions among magnetic and nonmagnetic states. The ground states of unstrained TaS$_2$ and TaSe$_2$ are Pauli paramagnetic, but a positive isotropic strain of 6% causes both TaS$_2$ and TaSe$_2$ to become ferromagnetic. Unstrained TaTe$_2$ exhibits a complicated spin structure with ze-ro magnetization, but the monolayer becomes ferromagnetic above a strain of 2%, with the magnetic moment of 0.28 $\mu_B$ per Ta atom. Hydrogenation changes the Ta$X_2$ monolayers from paramagnetic or antiferromagnetic metals to direct or indirect semiconductors close to metallic behavior.

Much theoretical and experimental work remains to be done on this intriguing class of transition-metal dichalcogenide thin films. However, since the switching between the involved magnetic and nonmagnetic states is realized relatively easily, we expect these materials to play an important role in the search for new materials for nanoelectronic and other applications.

## Acknowledgement

The authors are grateful to A. Kashyap for discussions relevant to the present paper. This work has been supported by ARO (W911NF-10-2-0099, P. M.) and DOE BES (DE-FG02-04ER46152, R. S., D. J. S.). Computations were performed at the University of Nebraska Holland Computing Center.